\documentclass[sigconf, 10pt, nonacm]{acmart}
\usepackage{xspace}
\usepackage{makecell}
\usepackage{threeparttable}
\usepackage{listings}
\usepackage{xcolor}
\usepackage{caption}
\usepackage{subcaption}
\usepackage{placeins}
\usepackage{float}
\usepackage{enumitem}
\usepackage{needspace}
\usepackage{nopageno}  
\usepackage{times}  
\usepackage{hyperref}
\usepackage{latexsym}
\usepackage[export]{adjustbox}
\usepackage{multirow}
\usepackage{tikz}
\usepackage{array}
\usepackage{comment}
\usepackage{url}
\usepackage{xurl}
\usepackage[normalem]{ulem}
\usepackage{mdwlist}
\usepackage{algorithm}
\usepackage{booktabs}
\usepackage[noend]{algpseudocode}
\usepackage{centernot}
\usepackage{fancyvrb}
\usepackage{alltt}
\usepackage[ampersand]{easylist}
\usepackage{fancybox}
\usepackage{graphicx}
\usepackage{tcolorbox}
\usepackage{balance}
\usepackage{xcolor}
\usepackage{verbatim}
\usepackage{breakurl}
\usepackage{xcolor}
\usepackage{listings}
\usepackage{framed}
\usepackage{lipsum} 
\usepackage[framemethod=tikz]{mdframed}

\newcommand{\para}[1]{\smallskip\noindent {\bf #1} }

\definecolor{codeblue}{rgb}{0,0,1}
\definecolor{codegreen}{rgb}{0,0.6,0}
\definecolor{codegray}{rgb}{0.5,0.5,0.5}
\definecolor{codepurple}{rgb}{0.58,0,0.82}
\definecolor{backcolour}{rgb}{0.95,0.95,0.92}
\definecolor{nocolor}{rgb}{1,1,1}

\definecolor{red}{rgb}{0.6,0,0} 
\definecolor{blue}{rgb}{0,0,0.6}
\definecolor{green}{rgb}{0,0.8,0}
\definecolor{cyan}{rgb}{0.0,0.6,0.6}
\definecolor{lightgray}{gray}{0.98}
\definecolor{lightblue}{rgb}{0.13, 0.67, 0.8}
\definecolor{lightorange}{RGB}{255,247,230}
\definecolor{codegreen}{rgb}{0,0.6,0}
\definecolor{codegray}{rgb}{0.5,0.5,0.5}
\definecolor{codepurple}{rgb}{0.58,0,0.82}
\definecolor{keywordcolor}{RGB}{94,20,64}
\definecolor{bluekeywords}{rgb}{0,0,1}
\definecolor{greencomments}{rgb}{0,0.5,0}
\definecolor{redstrings}{rgb}{0.64,0.08,0.08}
\definecolor{xmlcomments}{rgb}{0.5,0.5,0.5}
\definecolor{types}{rgb}{0.17,0.57,0.68}
\definecolor{KWColor}{RGB}{0,0,255}
\definecolor{AnnotationColor}{RGB}{0,137,180}
\definecolor{BlackColor}{RGB}{0,0,0}
\definecolor{CommentColor}{rgb}{0.12,0.38,0.18}
\definecolor{StringColor}{rgb}{0.06,0.10,0.98}
\definecolor{darkred}{rgb}{0.65,0,0}
\definecolor{lightgrey}{rgb}{0.8,0.8,0.8}
\definecolor{marmalade}{RGB}{193,101,18}
\definecolor{peach}{RGB}{250,217,193}
\definecolor{lime}{RGB}{220,237,193}
\definecolor{reqboxblue}{RGB}{28,35,105}
\definecolor{reqboxbg}{RGB}{248,248,255}

\newtcolorbox{requirementbox}{colback=reqboxbg,colframe=reqboxblue,boxrule=0.8pt,arc=1.2mm,left=4pt,right=4pt,top=3pt,bottom=3pt,before skip=0.6em,after skip=0.7em}

\lstdefinestyle{P4}{
  showspaces=false,
  showtabs=false,
  tabsize=2,
  columns=flexible,
  keepspaces=true,
  language={Java},
  numbers=left,
  xleftmargin=0pt,
  basicstyle=\ttfamily\footnotesize,
  commentstyle=\color{CommentColor}\ttfamily\footnotesize,
  stringstyle=\color{CommentColor},
  escapeinside={/*@}{@*/},
  numberstyle=\scriptsize\color{gray},
  showstringspaces=false,
  upquote=true,
  xleftmargin=1.2em,
  framexleftmargin=1.5em,
  keywords={ StateMachine }, 
  keywords=[2]{ bool, str, SM },
  keywords=[3]{ States, Transitions},   
  keywords=[4]{ read, assert, call, write},
  keywordstyle=\color{BlackColor}\bfseries,
  keywordstyle=[2]\color{codeblue},
  keywordstyle=[3]\color{red},
  keywordstyle=[4]\color{green},
  moredelim=[il][\color{darkgray}]{$$},
}

\mdfdefinestyle{background}{backgroundcolor=lightorange,innerrightmargin=0cm,innertopmargin=-0.1cm,innerbottommargin=-0.10cm,leftmargin=+0cm, roundcorner=2pt}

\mdfdefinestyle{tracebox}{
  backgroundcolor=lightorange,
  linecolor=black,
  linewidth=0.4pt,
  roundcorner=2pt,
  innertopmargin=0.55em,
  innerbottommargin=0.55em,
  innerleftmargin=0.8em,
  innerrightmargin=1.5em,
  skipabove=0pt,
  skipbelow=0pt
}

\usepackage[framemethod=tikz]{mdframed}

\usepackage[textsize=tiny,textwidth=0.6in]{todonotes}
\newcommand{\sys}{\textsc{CloudWeaver}\xspace}

\makeatletter
\renewcommand\@formatdoi[1]{\ignorespaces}
\makeatother

\makeatletter
\def\@copyrightspace{\relax}
\makeatother

\AtBeginDocument{%
  }

\setcopyright{none}
\acmISBN{}
\acmYear{}
\acmDOI{}

\newcounter{defn}[section]
\renewcommand{\thedefn}{\arabic{section}.\arabic{defn}}

\title[]{\huge{Towards a Systems Foundation for Agentic Cloud Management}}  

\begin{abstract}
Agentic cloud management is emerging as a practice to automate laborious operations,
    minimize toil, and improve responsiveness. 
Despite the rapid development of autonomous management agents,
    we argue that the fundamental missing piece
    is a systems foundation to enable 
    safe, effective operations across agents 
    and between agents and human operators. 
In this paper, we advocate for the need of such a systems foundation and share our efforts on developing \sys, an agentic management substrate
    that works across existing  cloud-user interfaces and future agent-native interfaces. 
Specifically, we discuss how \sys (1) scopes the context of individual agent sessions with local views of cloud resources and (2) coordinates concurrent management operations on shared cloud resources.
\sys offers strong safety guarantees and attributable feedback in the presence of conflicting intents, while preserving concurrency between independent operations.
We validate \sys using a representative Azure API workload.
\vspace{-5pt}
\end{abstract}

\author{%
Minghao Li$^{1}$,
Ziqian Liu$^{1}$,
Ziyu Mao$^{1}$,
Daqian Ding$^{1}$,
Yu Kang$^{2}$,
Qingwei Lin$^{2}$,
Tianyin Xu$^{3}$,
Yiming Qiu$^{1}$
}

\affiliation{%
  \institution{$^{1}$The University of Hong Kong \quad
  $^{2}$Microsoft \quad
  $^{3}$University of Illinois Urbana-Champaign}
  \country{}
}

\begin{document}

\maketitle







\section{Introduction}
Cloud 
infrastructures form the backbone of modern IT, but are notoriously hard to manage due to 
the fundamental split between the cloud providers and their tenants~\cite{qiu2023simplifying}.
The tenants manage cloud resources indirectly through multi-modal management interfaces across both low-level management actions (e.g., SDK or CLI~\cite{coleman2022cloud}) and higher-level Infrastructure-as-Code (IaC) workflows~\cite{terraform,pulumi}. 
The practice enables 
tenants to control complex cloud resources without owning the underlying datacenter infrastructures. 
However, this abstraction breaks at the boundary between scoped user-management intent and shared cloud infrastructure. 
A tenant often means parties with distinct resource domains, operational roles, and ownerships~\cite{artac2017devops}: one team may repair network topology, another auto-scale compute capacity, and another configure security policies. 
Each party acts through a scoped stream of observations and operations, without a complete and authoritative view of the tenant’s shared state. Once these operations reach the provider backend, their scope is lost: they become individual API calls whose effects unfold over shared resources~\cite{restler}. The caller retains the intent but not the feasibility of intent execution, while the provider sees the shared state but not the context that relates execution to the intent. Human operators bridge this gap through workflow orchestration, change windows, and manual recovery, leaving coordination as external conventions rather than a guarantee of the cloud-user interface.

Agentic cloud management now faces a fundamental choice: should AI agents be expected to reconstruct the missing coordination between management intent and cloud execution through reasoning, or should the cloud management stack provide it as a systems abstraction? We argue for the latter. 
Better AI may formulate stronger plans and recover from more failures, but they still reason from the context and feedback exposed by the system~\cite{swe-agent,aiopslab,aiops_1,aiops_2}. Asking probabilistic AI agents to reconcile conflicting intents by inferring session scope, concurrent effects, and execution outcomes from incomplete state and ambiguous feedback only compounds uncertainty. Coordination should therefore shift away from external best practices and be built directly into the cloud management abstraction itself.

In this paper, we advocate for a systems foundation to
enable agentic cloud management. 
Specifically, we envision a transparent \textit{coordination contract} beneath existing cloud-user interfaces and future agent-native interfaces, which offers the execution semantics required for reliable closed-loop agent control.
The contract binds each session’s context and actions to authoritative shared state, commits their effects through semantic transactions, and returns attributable feedback that renews the context of affected sessions. 
Realizing this vision would bridge the gap between scoped intent and shared state, making their correspondence a first-class systems invariant.

To obtain coherent reasoning context, the contract needs to enforce strong isolation across sessions.
This is nontrivial because each session-scoped view is a live execution status that must be constructed online, not from a snapshot~\cite{nsync}.
Furthermore, to be transparent to interface choices, coordination must sit at the boundary of provider APIs, where intents arrive as low-level operation streams. The session view must then play two important roles: (1) before execution, it checks admissibility: each incoming operation must be checked against the session’s current context to decide whether it can extend the view with valid resources, dependencies, and ownership, and (2) after admission, it tracks feedback: the same view must track pending, committed, failed, and shared effects so that the session can trust the observation. If either role is wrong, the session may reason about dependencies that never existed or act on resources it does not own. 

Safe coordination of concurrent operations is another challenge. A locally valid intent may not be globally executable as 
operations admitted by different sessions may still conflict with each other. 
The interface therefore needs an authoritative global view: a single source of truth that records resources, dependencies, and operations across all sessions. At first glance, concurrency control over this view may seem as simple as enforcing mutual exclusion on each cloud resource. In reality, it must reason about semantic effects: deleting a resource may invalidate another session’s dependency;
updating a parent resource may block child-resource creation;
and quota reservations may make otherwise independent intents incompatible. 
These effects 
determine which operations can run together, which must wait, and which should be denied before reaching the provider. Thus, concurrency requires a unified 
mechanism that makes global admission decisions with semantic-aware execution policies.

To address these challenges, we propose to build coordination atop three architectural pillars. 
First, the system must preserve management intent as it is lowered into provider API calls. It interprets each session’s API stream as semantic deltas over cloud resources, dependencies, lifecycle state, and capacity, thereby recovering the structure that conventional interfaces discard at the provider boundary.
Second, session-local and global views are coupled by a bidirectional projection: every local object and pending effect has a global counterpart, while changes in authoritative shared state are reflected back into the session context. This projection makes the correspondence between reasoning and execution a continuously maintained invariant rather than a condition repaired after drifts.
Third, concurrency control follows the semantics of each semantic delta. The transaction layer acquires structural rights for topology and lifecycle effects and reserves consumable rights for capacity and quota, allowing independent transitions to proceed while delaying those whose effects truly conflict.
Together, these principles preserve intent across interface lowering, ground every session in shared execution, and commit only globally valid cloud-state transitions.

We are building a substrate to realize the envisioned systems foundation, named \sys.
We discuss its core design with preliminary results, and discuss how it could enable future agentic cloud management endeavors.

\vspace{-5pt}

\section{Background}

{\bf Cloud Infrastructure Management.}
The management is exposed through a variety of cloud-user interfaces. 
Provider APIs define the authoritative operations over cloud resources. Most users invoke APIs indirectly: SDKs package them into libraries~\cite{bisong2019overview}, CLIs expose them as commands~\cite{shah2019building}, and web consoles support ClickOps with visual context~\cite{cloud_agent}. Above them, Infrastructure-as-Code (IaC) has become the dominant structured approach~\cite{qiu2023simplifying,peng2025automated}. IaC systems such as Terraform~\cite{terraform} and Pulumi~\cite{pulumi} let users specify desired cloud state, compare it against known state, construct an execution plan, and schedule the corresponding API operations. Despite their different abstractions, these interfaces ultimately converge at the same boundary: they translate tenant management intent into provider operations over shared cloud state.

\begin{figure}[t]
  \centering
  \includegraphics[width=0.48\textwidth]{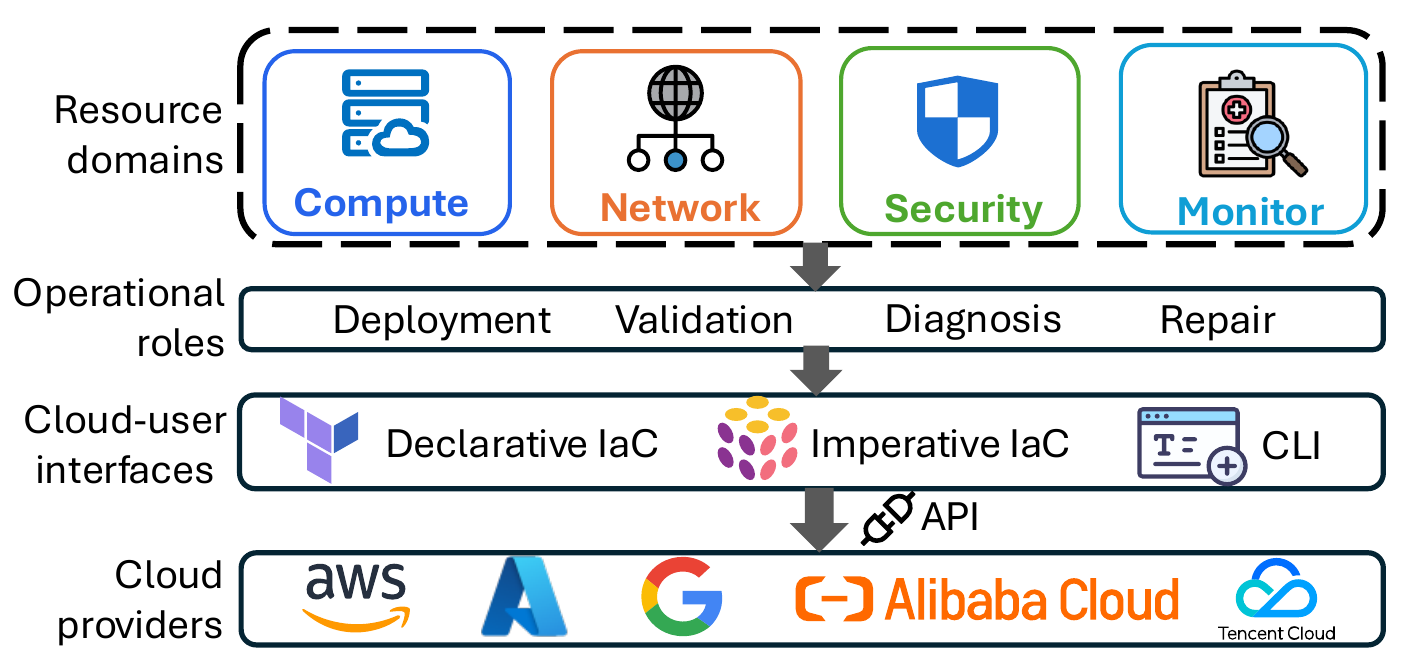}
  \vspace{-6mm}
  \caption{Modern cloud management stack.}
  \vspace{-6mm}
  \label{fig:agentic cloud}
\end{figure}

This stack has been increasingly used by AI agents. Platforms 
are exposing infrastructure state, operation context, and actions to agents. Providers such as Azure, AWS, and GCP now offer agentic services for operating and troubleshooting cloud systems~\cite{aws_services,gcp-gemini,azure-copilot}, while IaC ecosystems are introducing MCP interfaces that connect LLMs to infrastructure metadata and configuration workflows~\cite{pulumi-internals,pulumi-resource-providers}. 
Recent AI benchmarks are also moving towards system operations and infrastructure management~\cite{clark2026sregym,cloud-bench1,iac-eval,itbench}. Together, these efforts point to a new paradigm in which agents become active participants in the observe–plan–execute loop, acting over the same provider state as human operators~\cite{chen:stratus:2025,aiops_2,guo2024owl,liu2025large,shetty2024building,yang2026tsguard}.

Figure~\ref{fig:agentic cloud} depicts the landscape of modern cloud management stack. Within a tenant, responsibility is commonly divided across resource domains (e.g., compute, network, security, and monitoring) and across operational roles (e.g., deployment, diagnosis, and repair). Each party operates from a scoped management context, while its actions ultimately affect infrastructure shared with other parties. As agents join these workflows, cloud management increasingly consists of multiple human operators and agent sessions independently observing, planning, and acting over the shared provider state.

\para{State of the Art \& Limitations.} Existing cloud management lacks 
a coordination system---management operations are directly submitted to the provider and rely on the cloud backend to reject invalid ones. In other words, it treats provider errors as a coordination mechanism, but provider errors are late feedback~\cite{gu2023acto,sun2024anvil}. By the time an operation fails, it has already been interpreted by complex provider internals, and the caller must infer whether the cause was stale context, a hidden dependency, exhausted quota, an in-flight provider conflict, etc. The ambiguity is painful for humans and toxic for agents whose next action is conditioned on the feedback. 

The opposite extreme is to impose a single global lock on all management operations. 
However, most operations do not conflict, and existing IaC systems are explicitly designed to exploit concurrency among independent changes~\cite{qiu2024unearthing}. Terraform~\cite{terraform}, for example, constructs a dependency graph and executes non-dependent operations in parallel. Serialize-all is therefore not an appropriate abstraction either.

While IaC exposes parallelism, it is confined to the state within a workflow; concurrent changes issued by other workflows remain outside its view. Even within this boundary, IaC does not fully capture execution conflicts~\cite{terraform-lifecycle,terraform-revert}. IaC provider plugins may rely on coarse, resource-specific mutexes to serialize operations; such locks approximate conflicts, without fine-grained coordination. IaC drift-reconciliation systems~\cite{snyk-drift-tools,terraform-drifts,pulumi-drifts} like NSync~\cite{nsync} propagate out-of-band changes back into the IaC program, but only after its view has diverged from the cloud. The limitation is therefore not drift itself, but a workflow-scoped execution model that is incomplete within its boundary and blind beyond it.

Modern cloud orchestration platforms often partition infrastructure into independently managed units and workflows, with HCP workspaces~\cite{terraform-workspaces} and Spacelift stacks~\cite{spacelift} as two common examples. These boundaries are useful for organizing state, ownership, and execution, but they do not determine whether concurrently issued changes can safely coexist over shared provider state. Conflicts may cross partitions through shared dependencies or capacity, while distinct intents within the same partition may still overlap during execution. This mismatch becomes especially problematic for agentic management, where sessions are often created around short-lived intents rather than durable administrative boundaries. 

\vspace{-2mm}
\section{Towards a Systems Foundation}

\subsection{A Motivating Example}
\begin{figure}[t]
  \centering
  \includegraphics[width=0.48\textwidth]{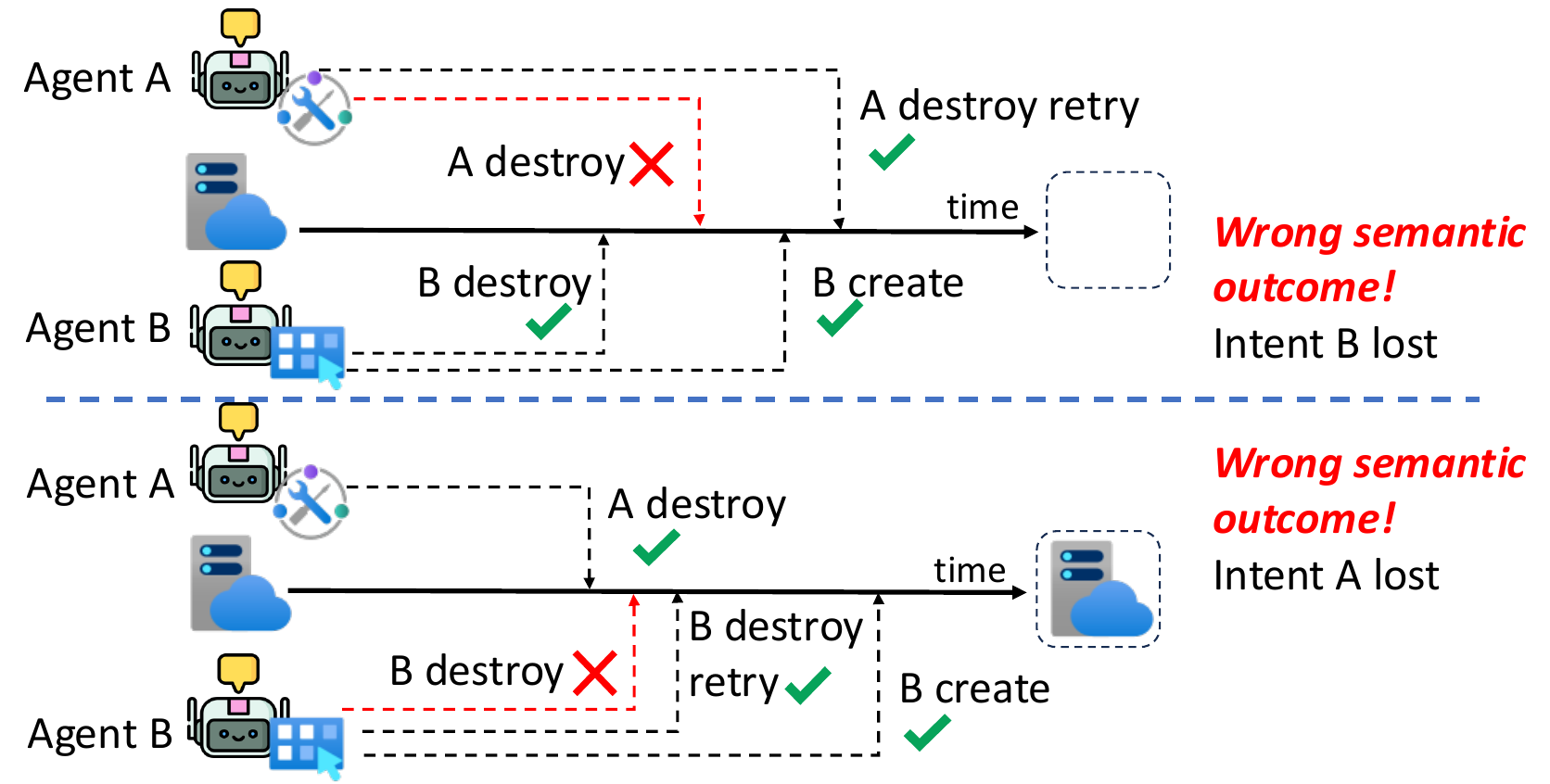}
  \vspace{-7mm}
  \caption{A motivating intent conflict example.}
  \vspace{-4mm}
  \label{fig:3.1}
\end{figure}

Figure~\ref{fig:3.1} presents a scenario involving two agents operating on the same VM. Agent~A intends to permanently destroy the VM, while Agent~B concurrently intends to replace it through a destroy–then–recreate workflow. If Agent~B’s destroy executes first, Agent~A’s subsequent destroy fails because the VM no longer exists. Agent~B then recreates the VM, while Agent~A continues retrying until it destroys the replacement, causing Agent~B’s intent to be lost. Conversely, if Agent~A’s destroy executes first, Agent~B’s destroy fails. Agent~B then observes that the VM has already been removed and proceeds to recreate it, causing Agent~A’s intent to be lost.

This case exposes two distinct limitations. First, the final outcome depends on operation timing and retry order. Without concurrency control over active intents, neither agent nor the system determines which intent should prevail; that decision is effectively delegated to incidental API interleavings and independent retry loops. This makes execution nondeterministic from the agents’ perspective. Second, the feedback available to agents lacks intent attribution. The first rejection reveals only an operational conflict, not that another session is pursuing an incompatible intent. 
If agents had learned the conflicting intent with the first rejection, they could have revised their plans instead of blindly retrying. What is missing is a system-level abstraction that connects session-scoped management intent to authoritative shared state throughout planning, execution, and feedback.

\vspace{-1mm}
\subsection{Design Principles}
\label{sec:vision}

We view coordination as a common foundation for agentic cloud management. By grounding each intent in session-scoped context and coordinating concurrent operations over shared infrastructure, this foundation provides reliable execution semantics upon which new agent-native interfaces, validation and recovery mechanisms, agent architectures, and other management capabilities can be built. We distill this foundation into two design principles as shown below.
 
\para{Session-scoped cloud views.} Cloud-management agents operate on session-scoped intents, but realize those intents over resources and constraints shared across sessions. Therefore, a session boundary must be defined as a policy-controlled projection over reconciled global state, rather than as a private physical copy. This projection may mix private resources and explicitly shared resources, while the underlying global infrastructure should remain hidden. 
If shared global state is exposed directly, the session abstraction breaks: agents may observe unrelated resources, cleanup artifacts, failures, or transient states; observations become unstable under concurrency; and the system cannot enforce session-level ownership, visibility, or access permissions. Raw global state also lacks the session-level topology, dependencies and recent operational context needed for planning and verification. Leakage appears during execution: one session may fail because another consumes shared capacity, triggers rate limits, or mutates shared parents. Such effects may look intent-relevant to the agent but are actually uncontrolled cross-session interference.

\begin{requirementbox}
\textbf{Principle 1.} Agentic cloud management must maintain session-scoped local views grounded in global state, with bounded observation and action space.

\end{requirementbox}

\para{Coordination under semantic constraints.}
Cloud coordination must reason about the semantic scope and effects of operations rather than resource identity alone. Many conflicts arise from identifier, lifecycle, and configuration semantics, such as scoped uniqueness, in-flight resource mutations, and compatibility constraints across related resources.
For example, if one agent is updating a VM, another agent should not concurrently update the same VM, resize its attached disks, or attach or detach NICs, because these operations interact through the VM's lifecycle and attachment state.
Other conflicts are capacity-related, including user-defined limits such as shared-disk attachment limits and subnet usable-IP capacity, as well as provider-defined limits such as regional VM-family vCPU quota or NAT/public-IP outbound-port quota.
Capacity constraints are also semantic rather than simple counters. For example, a shared data disk with \texttt{maxShares=2} exposes two attachment slots, but this capacity is scoped to that particular disk and is consumed only by operations that attach the same disk.
Coordination must therefore identify semantic conflicts before provider execution, preventing delayed failures, leaked resources, and cross-session interference while preserving parallelism among compatible operations.

\begin{requirementbox}
\textbf{Principle 2.} Agentic cloud management must enforce semantic constraints before provider execution, preventing unsafe side effects while preserving safe concurrency.
\end{requirementbox}

\begin{figure}[t]
  \centering
  \includegraphics[width=0.44\textwidth]{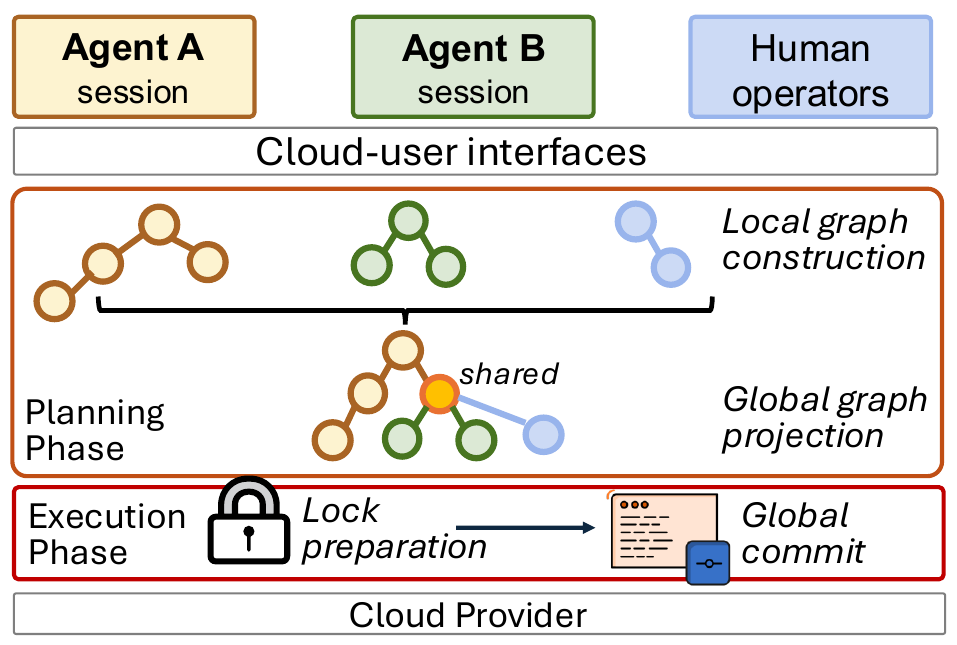}
  
  \vspace{-4mm}
  \caption{\sys overview.}
  \vspace{-4mm}
  \label{fig:workflow overview}
\end{figure}

\section{\sys}

We are developing \sys, a system substrate that offers safe and efficient agentic cloud management. Figure~\ref{fig:workflow overview} gives an overview of \sys. \sys separates planning from execution to realize the two principles discussed in \S\ref{sec:vision}. For Principle~1, \sys's planning phase maintains local and global views of shared cloud state---local views expose each agent to an isolated and session-scoped environment, while the global view captures underlying shared resources and cross-session dependencies. For Principle~2, \sys's execution phase coordinates operations before provider submission using structural and capacity-aware synchronization. 
In this section, we present the two key components of \sys.

\begin{figure}[t]
  \centering
  \includegraphics[width=0.49\textwidth]{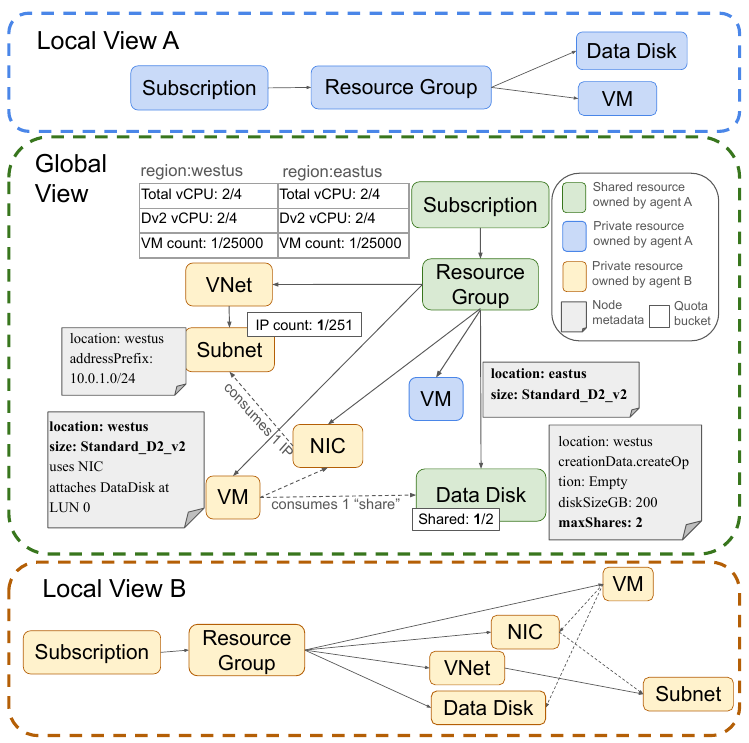}
  \vspace{-5mm}
  \caption{Global and local views of shared resources.}
  \vspace{-5mm}
  \label{fig:control-plane}
\end{figure}

\subsection{Session Management for Agent Planning}

To support the planning phase of AI agents, \sys maintains shared cloud state in two representations: a global graph and local graphs.
The global graph $G^{\mathrm{g}}=(V^{\mathrm{g}},E^{\mathrm{g}})$ records the reconciled shared state across sessions.
Each vertex in $V^{\mathrm{g}}$ represents a cloud resource such as a VM, a NIC, a subnet, or a disk, while each edge in $E^{\mathrm{g}}$ captures semantic relations such as containment, attachment, routing, reference, or dependency. Vertices are annotated with provider identifiers, lifecycle states, ownership and visibility labels, and intent history. Intent history helps agents identify why and how the resource transformed to the current state as well as who initiated the operations. In general, the global graph serves as the coordination substrate that records what physically exists, what is shared or reserved, and what constraints must be considered before provider execution.

For each session $i$, \sys exposes a local graph $G_i^{\mathrm{l}}$, a policy-controlled projection of $G^{\mathrm{g}}$.
The local view contains private and explicitly shared resources visible to the agent. Local resources and relations are mapped from their global counterparts by \sys, allowing the same shared global resource to appear in multiple sessions without revealing the underlying physical sharing. Figure~\ref{fig:control-plane} illustrates this projection. The global graph records both topology and constraints: a VM in \textit{westus} consumes regional vCPU quota, its NIC consumes subnet address capacity, and a shared disk consumes one \textit{maxShares} slot. The projection policy exposes shared resources in the relevant local views while keeping private resources visible only to their owner sessions.

At runtime, the local view is constructed and maintained through two separate paths. First, the API parser processes each incoming request and matches each to a provider plugin rule. It evaluates the rule against the current \(G_i^{\mathrm{l}}\), and performs a preflight condition check on whether the request is valid under the session's visible resources, dependencies, guards, and capacity constraints. If valid, the parser derives a local graph delta \(\Delta G_i^{\mathrm{l}}\), including resource updates, semantic edge modifications, extracted metadata, and constraint effects. The projection layer then resolves local resource identifiers from the delta graph to produce a global graph delta \(\Delta G^{\mathrm{g}}\), which exposes the affected resources, relations, and constraints to the execution phase for coordinated provider execution. After execution completes, the cloud provider responses and reconciliation results are fed back to update both \(G^{\mathrm{g}}\) and each affected \(G_i^{\mathrm{l}}\). This feedback path captures asynchronous completion, failures, and drift to ensure that later admission checks are performed against a safe and up-to-date local view rather than a stale session snapshot.

\subsection{Semantic Transaction for Agent Execution}

\begin{table}[t]
\centering
\small
\caption{MGL modes for cloud transactions.}
\vspace{-2mm}
\label{tab:mgl-locks}
\begin{tabular}{ll}
\toprule
Mode & Meaning \\
\midrule
\(X\)  & Exclusive access to a node and its descendants. \\
\(S\)  & Shared access to a node and its descendants. \\
\(IX\) & Intention to acquire exclusive locks below this node. \\
\(IS\) & Intention to acquire shared locks below this node. \\
\(NL\) & No lock. \\
\bottomrule
\end{tabular}
\vspace{-4mm}
\end{table}

A \sys transaction is an internal execution unit derived from an intercepted cloud operation. It binds the issuing session and its management intent to the projected global delta of the operation $\Delta G^{\mathrm{g}}$, the structural lock requirements, the affected capacity constraints, and the execution status. The transaction allows \sys to determine whether the operation can be safely admitted before forwarding the corresponding request to the provider. To coordinate these transactions, \sys uses Multi-Granularity Locks (MGL) to coordinate operations across hierarchical resource dependencies, and Escrow Locks to reserve capacity without serializing independent transactions.

MGL protects the resource topology encoded in $\Delta G^{\mathrm{g}}$, including containment, attachment, and dependency relations. For each transaction $T$, the runtime derives a structural lock set $L_T$ over the touched nodes and edges: modified resources receive exclusive locks, while their ancestors receive intention locks. For example, creating a VM may acquire $IX$ locks on its parent region, VNet, and subnet, and $X$ locks on the new VM and NIC; deleting the subnet would require an $X$ lock on the subnet and therefore conflict with this child-resource creation.  Table~\ref{tab:mgl-locks} summarizes the MGL modes. 

Escrow locks protect capacity constraints such as regional quota, subnet IP capacity, and shared-disk attachment slots. Each capacity bucket tracks reserved capacity bounds, and an operation can be admitted only if its effect keeps the bucket feasible. This allows concurrent operations to consume disjoint capacity requests while preventing overbooking.

\begin{figure}[t]
  \centering
  \includegraphics[width=0.49\textwidth]{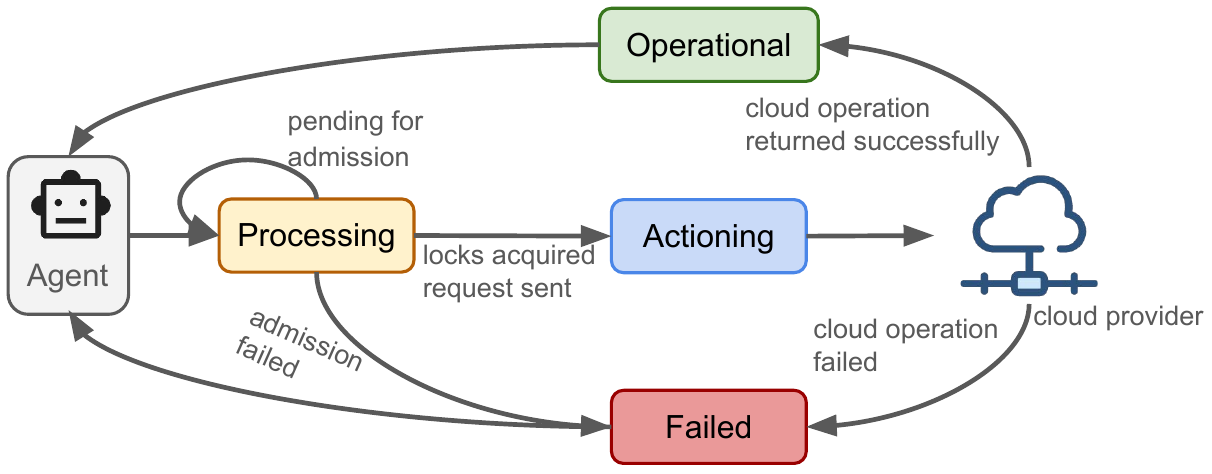}
  \vspace{-4mm}
  \caption{Transaction state transitions.}
  \vspace{-5mm}
  \label{fig:state-machine}
\end{figure}

Execution proceeds in two steps to avoid holding expensive locks for a long duration. Figure~\ref{fig:state-machine} shows the lifecycle of a transaction. In the planning phase, an intercepted request is parsed into a transaction descriptor containing a lock plan derived from a snapshot of the global graph with status \textit{Processing}. This computation runs outside the global critical section, allowing independent sessions to plan concurrently. In the commit phase, the runtime revalidates the descriptor against the current global graph, acquires the required MGLs, and reserves escrow capacity. If admission fails, the request is not sent to the provider and is rejected, queued, or retried according to policy. Alongside the decision, \sys returns an attributable conflict report identifying the conflicting resources and constraints together with the intent of the transaction currently blocking admission. If admission succeeds, the admitted delta is installed in $G^{\mathrm{g}}$ with status \textit{Actioning}, local--global mappings are recorded, and the API request is forwarded. Upon completion, the runtime reconciles the provider response with the graph: a successful operation transitions the affected resources to \textit{Operational}, whereas an unsuccessful operation marks them as \textit{Failed} and releases or rolls back the associated reservations.

\section{Preliminary Validation}

We validate \sys using an Azure trace, showing that \sys preserves concurrency among independent operations while preventing semantic conflicts.

\begin{figure}[t]
  \centering
    \includegraphics[width=0.9\linewidth]{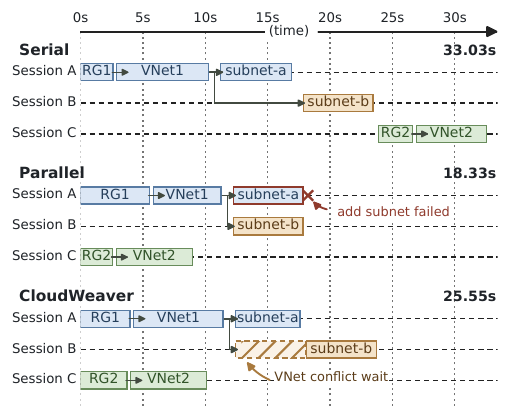}
    \vspace{-5mm}
    \caption{Timeline of the preliminary validation trace.}
    \label{fig:primary-validation-timeline}
  \vspace{-5mm}
\end{figure}

\para{Experimental workload trace.} The workload contains six Azure API operations from three sessions. Session A creates a shared resource group and VNet, while Session C creates an independent resource group and VNet. This independent branch represents safe parallelism that should not be serialized with A's work. After creating the shared VNet, Sessions A and B add two distinct subnets to it. The API sequence treats these requests as independent because they target different subnet resources, allowing them to become ready at the same time. However, both operations mutate the same Azure VNet parent and therefore cannot safely execute concurrently under Azure’s provider semantics.

\para{Execution behavior.} We compare \sys against two execution policies. The \textit{serial} policy executes one operation at a time across sessions. This avoids provider conflicts, but also serializes independent work such as Session C's VNet creation. 
The \textit{parallel} policy executes all dependency-ready operations concurrently, overlapping independent branches while issuing both subnet writes to the shared VNet at the same time. 
In our run, Azure accepts one request and rejects the other with \texttt{HTTP 409} and \texttt{AnotherOpInProgress}, because an operation on the same VNet parent is still in progress.
This failure shows that the \textit{parallel} policy understands explicit request ordering, but not provider-level conflict scopes, causing it to issue two distinct subnet requests concurrently even though Azure serializes both at their shared VNet parent.
In contrast, \sys supports semantic coordination, and admits API sequences through the transaction layer before forwarding them to Azure.

\para{Latency comparison.} Figure~\ref{fig:primary-validation-timeline} shows that serial execution completes all six operations in 33.03s, while unconstrained parallel execution finishes in 18.33s but succeeds on only five operations. The interface completes all operations in 25.55s, reducing makespan by 7.48s (22.6\%) while preserving correctness. It overlaps Session C's independent operations with Session A's shared-resource setup, while serializing conflicting subnet writes at the VNet scope. This preserves safe parallelism and avoids provider-side failures.

\section{Discussion and Future Work}
\para{Scalability and safety.}
\sys isolates provider-specific semantics in plugins, allowing each provider to define rules without modifying the coordination protocol. Developing these plugins may be largely automated by deriving cloud-operation semantics from API specifications, documentation, and traces~\cite{emu1,localstack,moto,azurite}, which we leave as future work. The deployment overhead of \sys is modest because the system operates only on compact resource metadata, dependency edges, lock states, and escrow counters. Its compute cost therefore depends mainly on the scope of each operation rather than the total cloud size. 
To ensure the safety of our semantic transaction protocol, we further validated it with TLA+, using TLC for bounded safety and progress checking and TLAPS for a general deadlock-freedom proof.

\para{Intent-aware agentic interface.}
An agent-native interface built on \sys should be organized around management intents rather than individual provider API calls or global updates. It would let an agent declare an objective, inspect the scope of resources and actions available to its session, submit a multi-step plan for validation, and receive structured outcomes that distinguish admissible actions, deferred actions, and conflicts requiring replanning. Rather than exposing provider-specific failures as unstructured text, the interface should return machine-consumable explanations and support explicit follow-up actions such as revise, wait, abort, or verify. Such an interface would make closed-loop cloud management a first-class interaction model, while \sys supplies the underlying state and coordination needed to implement it.

\para{Implications for RL training.}
\sys enables scalable RL training on real cloud infrastructure by allowing many agent sessions to share warm resources while each observes a clean, session-scoped environment. Before each training batch, the system can determine which sessions can reuse existing accounts, networks, images, and pre-created resources based on task requirements, provisioning latency, quota pressure, and isolation constraints. View isolation prevents cross-session interference from corrupting trajectories and rewards, while semantic coordination preserves safe parallelism. Together, they reduce setup cost and latency, improve rollout throughput, and retain real-provider fidelity.

\section{Concluding Remarks}
Agentic cloud management requires more than capable AI agents: it requires a systems foundation that keeps their reasoning grounded in the cloud infrastructure that they jointly change. This paper discusses design principles for such a coordination substrate, and presents \sys as an initial realization. 
Looking forward, we envision this foundation supporting a broader evolution of agentic cloud management, from today’s copilots toward persistent, collaborative, and increasingly autonomous systems for planning, validation, recovery, and operation.

\bibliographystyle{abbrv} 
\bibliography{citation}

\end{document}